\documentclass[preprint,amsmath,amssymb,a4paper,nofootinbib,superscriptaddress,floatfix]{revtex4-1}
\usepackage{graphicx}
\usepackage{natbib}
\usepackage{dcolumn}
\usepackage{bm}
\DeclareMathOperator{\sech}{sech}

\makeatother
\newcommand{\Glasgow}{SUPA, School of Physics and Astronomy, University of Glasgow, Glasgow G12 8QQ, UK}

\newcommand{\ParisSud}{Institut d'Electronique Fondamentale, Univ. Paris-Sud, 91405 Orsay, France}
\newcommand{\CNRS}{UMR 8622, CNRS, 91405 Orsay, France}

\begin{document}

%
%
\title{Spin Wave Eigenmodes of Dzyaloshinskii Domain Walls}

\author{Pablo Borys}
\affiliation{\Glasgow}
\author{Felipe Garcia-Sanchez}
\affiliation{\ParisSud}
\affiliation{\CNRS}
\author{Joo-Von Kim}
\affiliation{\ParisSud}
\affiliation{\CNRS}
\author{Robert L. Stamps}
\affiliation{\Glasgow}


\date{\today}

\begin{abstract}

A theory for the spin wave eigenmodes of a Dzyaloshinskii domain wall is presented. These walls are Neel-type domain walls that can appear in perpendicularly-magnetized ultrathin ferromagnets in the presence of a sizeable Dzyaloshinskii-Moriya interaction. The mode frequencies for spin waves propagating parallel and perpendicular to the domain wall are computed using a continuum approximation. In contrast to Bloch-type walls, it is found that the spin wave potential associated with Dzyaloshinskii domain walls is not reflectionless, which leads to a finite scattering cross-section for interactions between spin waves and domain walls. A gap produced by the Dzyaloshinskii interaction emerges, and consequences for spin wave driven domain wall motion and band structures arising from periodic wall arrays are discussed.
\end{abstract}
\maketitle
\section{Introduction}
The Dzyaloshinskii-Moriya interaction (DMI) is an antisymmetric contribution to the exchange energy that can exist in spin systems that lack inversion symmetry.\cite{dzyaloshinsky_thermodynamic_1958, moriya_anisotropic_1960, moriya_new_1960} Spin textures stabilized by DMI are of special interest due to the possibility of new technological applications.\cite{fert_skyrmions_2013, PhysRevB.88.184422, brataas_spintronics:_2013} For room temperature operation, interface DMI is of particular importance in terms of thin film structures based on transition metals, compatible with traditional spintronic devices. It has been shown in perpendicular materials that because of DMI compensation of the dipole-dipole interaction at the center of a domain wall, a N{\'e}el-type domain wall is favored  with important enhancements of domain wall stability and mobility.\cite{thiaville_dynamics_2012, tetienne_nature_2015} 

Small amplitude fluctuations of the magnetisation about equilibrium-- spin waves-- in systems with DMI have been studied experimentally, \cite{zakeri_asymmetric_2010, udvardi_chiral_2009} and theoretically\cite{moon_spin-wave_2013, PhysRevB.89.224408} for interface DMI films. A key feature is nonreciprocity of frequency as a function of propagation direction for finite wavelength spin waves, which emerges from the chiral symmetry breaking DMI.  In the present work we examine the spin wave dispersion in a film containing a DMI stabilized N{\'e}el wall. We find that spin waves are partially reflected by the domain wall due to an extra chiral term in the potential associated with the domain wall. Local modes are found, and we show that DMI drives a hybridisation of traveling spin waves with these localized states. 

The hybridisation produces an energy-split dispersion with a gap magnitude that is proportional to the magnitude of the DMI. We illustrate this effect with a suggestion for a magnonic crystal produced by a periodic array of domain walls in a nanowire with DMI. Without DMI the walls are Bloch-type walls which are known to represent reflectionless potentials for spin waves traveling through them.\cite{winter_bloch_1961, braun_fluctuations_1994, bayer_phase_2005} This results in a gapless band structure. However, with DMI the stable configuration becomes an array of N{\'e}el-type walls with a modified potential for the spin waves. The modified potential is no longer reflectionless and standing waves appear at the edges of the first Brillouin zone producing gaps in the band structure.

This article is organized as follows. In Section II, the model and calculations involving the static domain wall profile are presented. In Section III, the spin wave eigenmodes of the Dzyaloshinskii domain wall (DDW) are computed using a variational method in the continuum approximation. Consequences of these eigenmodes are then explored in Section IV, where the reflection and transmission coefficients for propagating spin waves through the domain wall are computed and band gaps in associated with periodic wall arrays are discussed. Finally, a discussion and some concluding remarks are given in Section V.

\section{Model and static wall profile}

\begin{figure}
\centering\includegraphics[width=6.0cm]{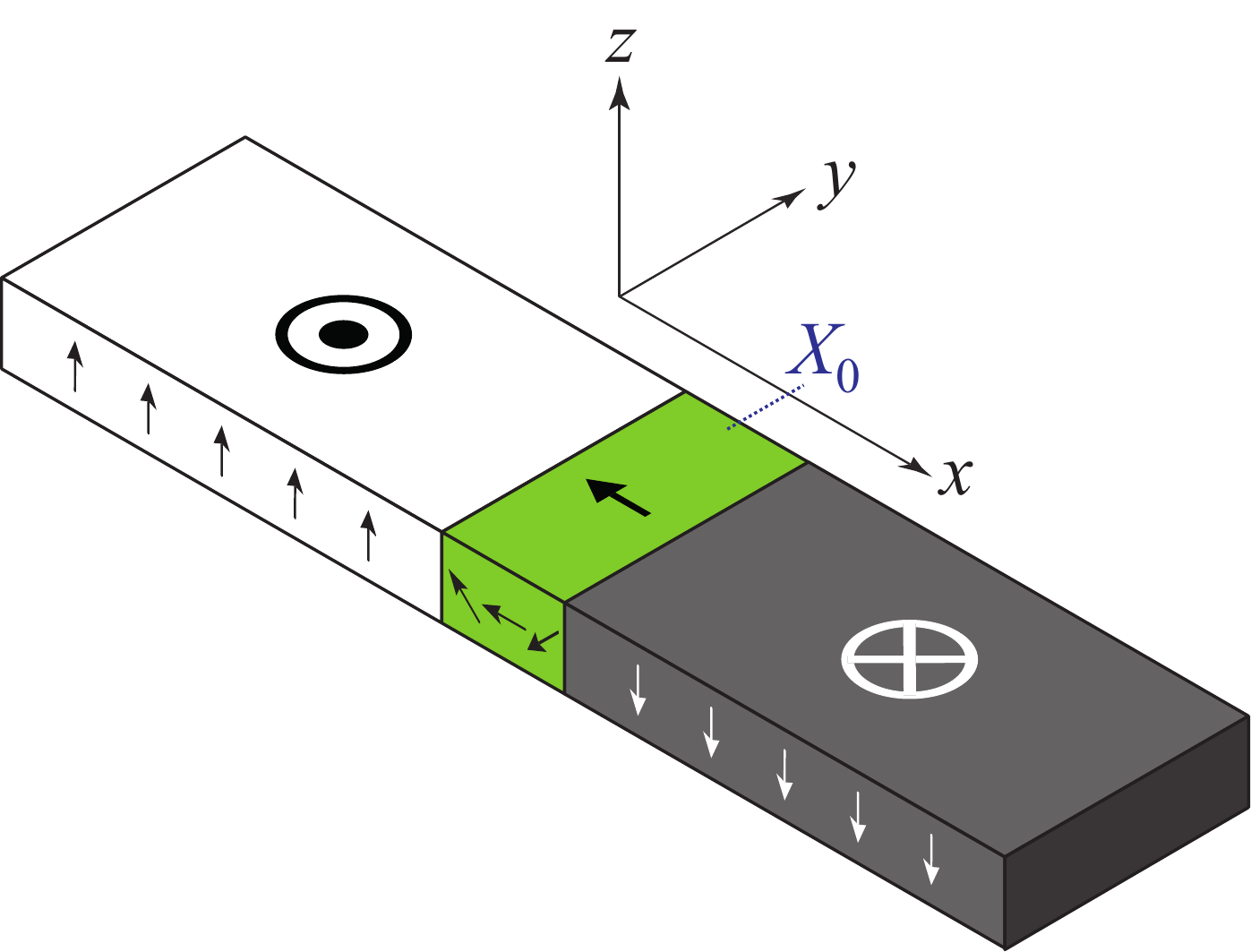}
\caption{(Color online) Geometry considered for the N{\'e}el-type Dzyaloshinskii domain wall. $X_0$ denotes the position of the wall center along the $x$ axis. Translational invariance is assumed along the $y$ direction and the magnetization is taken to be uniform across the thickness of the film in the $z$ direction.}
\label{fig:geometry}
\end{figure}

An ultrathin ferromagnetic wire is considered in which a domain wall separates two uniformly-magnetized domains along the $x$ axis, as shown in Figure (\ref{fig:geometry}). 
The uniaxial anisotropy, $K_{u}$, is taken to lie along the $z$ axis, perpendicular to the film plane, while a transverse anistropy resulting from volume dipolar charges, $K_\perp$, is present along the $x$ axis. In addition, an interfacial Dzyaloshinskii-Moriya interaction  $D$ is also considered, with a form consistent with a multilayered system with a heavy-metal subtrate.\cite{bogdanov_chiral_2001, thiaville_dynamics_2012} The form of this interaction can be written in terms of the Lifshitz invariants $L^k_{ij}=m_i\frac{\partial m_j}{\partial x_k}-m_j\frac{\partial m_i}{\partial x_k}$ as $D\,(L^x_{zx}+L^y_{zy})$. The magnetization orientation, represented by the unit vector $\mathbf{m}$, is parametrized using spherical coordinates as $\mathbf{m} = \left( \sin\theta \cos\phi, \sin\theta \sin\phi, \cos\theta \right)$, where $\theta = \theta(\mathbf{r},t)$ and $\phi = \phi(\mathbf{r},t)$.
The total magnetic energy of this system is given by the functional $U[\theta(\mathbf{r}),\phi(\mathbf{r})]$,
\begin{multline}
U = \int dV \; \left[ A \left( \left(\nabla \theta \right)^2 + \sin^2\theta \left(\nabla \phi \right)^2 \right) \right.\\
+ \left( K_u + K_\perp \cos^2\phi \right) \sin^2\theta \\
+ D \left( \frac{\partial \theta}{\partial x} \cos\phi + \frac{\partial \theta}{\partial y} \sin\phi \right. \\
\left. \left. + \frac{1}{2}\sin2\theta \left(\frac{\partial \phi}{\partial y} \cos\phi - \frac{\partial \phi}{\partial x} \sin\phi  \right) \right) \right],
\label{eq:energy}
\end{multline}
where $A$ is the exchange constant. The static profile of the domain wall is determined by the solution to the Euler-Lagrange equations associated with the functional in Equation (\ref{eq:energy}), which are obtained by setting the first-order functional derivatives to zero. By neglecting variations in the $y$ direction and assuming the solution $\phi(x) = \phi_0$, the nonlinear differential equations satisfied by the static wall profile $(\theta_0, \phi_0)$ are given by
\begin{align}
\frac{\delta U}{\delta \theta} &= 0 \Rightarrow A \frac{\partial^2 \theta_0}{\partial x^2} - \frac{1}{2}\left(K_u + K_\perp \cos^2(\phi_0) \right)\sin(2\theta_0) = 0, \\
\frac{\delta U}{\delta \phi} &= 0 \Rightarrow \sin(\phi_0) \sin^2(\theta_0) \left(K_\perp \cos(\phi_0) + D \frac{\partial \theta_0}{\partial x} \right) = 0.
\end{align}
Note that the second equation above is satisfied by the $\phi_0$ ansatz for finite values of the DMI, $D \neq 0$, only if the domain wall assumes a pure N{\'e}el profile ($\phi_0 = 0, \pi$). By assuming a N{\'e}el wall state, the solution for $\theta_0(x)$ to be written as
\begin{equation}
\theta_0(x) = 2 \tan^{-1} \left[ \exp\left(\pm \frac{x- X_0}{\lambda} \right) \right],
\end{equation}
where $\lambda = \sqrt{A/(K_u + K_\perp)}$ is the domain wall width parameter and $X_0$ denotes the wall center. The solution with the positive sign in the argument of the exponential function gives the configuration illustrated in Figure (\ref{fig:geometry}). With this solution, the total domain wall energy (\ref{eq:energy}) can be evaluated to be 
\begin{equation}
\sigma_w \equiv U[\theta_0,\phi_0] = 4\sqrt{A \left(K_u + K_\perp \right)} \mp \pi D,
\end{equation}
where the negative sign corresponds to the solution $\phi_0 = \pi$ and the positive sign to $\phi_0 = 0$, which indicates that left-handed N{\'e}el walls are preferred energetically for $D>D_c=4\lambda K_\perp/\pi>0$.
\section{Spin wave Hamiltonian}

The magnetic energy functional  can be expanded up to second order in small fluctuations $(\delta\theta,\delta\phi)$ around the stable configuration $(\theta_0,\phi_0)$ to obtain the spin wave Hamiltonian,\cite{braun_fluctuations_1994, le_maho_spin-wave_2009}
\begin{equation}\label{eqn:swHam}
\delta H=\frac{K_\perp}{\kappa}\int dx\;\;\delta\theta\, V_P(x)\,\delta\theta+\delta\phi\left[V_P(x)-\Delta(x)-\kappa\right]\delta\phi.
\end{equation}
The energy of the fluctuations is described by the operators $V_P(x)=[-\lambda^2\partial_x^2+1-2\sech^2(x/\lambda)]$, $\Delta(x)=(D\,\kappa/\lambda K_\perp)\sech(x/\lambda)$ and $\kappa=K_\perp/(K_u+K_\perp)$. The Schr\"{o}dinger type operator, $V_P(x)$, has been widely studied and is used to describe spin waves in a Bloch type domain wall.\cite{winter_bloch_1961, kishine_adiabatic_2010} Solutions to these operator include a single bound state,
\begin{equation}
\xi_{loc}(x)=\dfrac{1}{\sqrt{2\lambda}}\sech(x/\lambda),
\end{equation}
with zero corresponding energy, and continuum-traveling states, 
\begin{equation}
\xi_k(x)=\dfrac{1}{\sqrt{\omega_k}}e^{ikx}[\tanh(x/\lambda)-ik\lambda],
\end{equation}
with eigenenergy given by $\omega_k=1+k^2\lambda^2$. The above states  form a complete orthonormal set,
\begin{equation}
\begin{split}
\int\; dV\; \xi^*_k\xi_{loc} & =0,\\
\int\;dV\;\xi_k^*\xi_m & =\delta_{k,m}.
\end{split}
\end{equation} 
From Equation (\ref{eqn:swHam}) it can be deduced that $\kappa$ introduces a constant ellipticity in the precession of the fluctuations, and DMI introduces a spatially dependent one through the $\Delta(x)$ term so that it is not trivial to find a basis that diagonalizes the spin wave Hamiltonian. We propose a linear superposition of the local and traveling modes,
\begin{equation}
\chi(x)= \delta\phi(x)+i \delta\theta(x)=ic_{loc}\xi_{loc}(x)+\sum_k d_k\xi_k(x),
\end{equation} 
to calculate the spin wave energy. After the space integrals are computed, we find  
\begin{multline}\label{eqn:ampl}
\delta H=-c_{loc}^2\left(\frac{\pi D}{4\lambda}+K_\perp\right)+\sum_k\left[A_k'd_k^*d_k+B_k'(d_k^*d_{-k}^*+d_kd_{-k}\right.\\\left.+C_kc_{loc}(d_k+d_k^*)\right]+
\sum_{km}U_{km}\,d_k^*d_m+V_{km}(d_kd_m+d_k^*d_m^*),
\end{multline}
where the coefficients are given by
\begin{equation}
\begin{split}
A'_k&=\omega_k(K_u+K_\perp)-\frac{K_\perp}{2},\\
B'_k&=\frac{K_\perp}{4},\\
C_k&=\int\;dV\;\xi_{loc}\Delta(x)\,\xi_k,\\
U_{km}&=\int\;dV\;\xi_k^*\Delta(x)\,\xi_m,\\
V_{km}&=\int\;dV\;(\xi_k\Delta(x)\xi_m+\xi_k^*\Delta(x)\xi_m^*).\\
\end{split}
\end{equation}
The $A'_k$ and $B'_k$ terms denote elliptical spin precession as a result of the transverse anisotropy and correspond to the usual terms found in the Bloch wall case.\cite{le_maho_spin-wave_2009, garcia-sanchez_2015_PRL} The $C_k$, $U_{km}$ and $V_{km}$ terms are proportional to the strength of $D$ and depend on $k$ because these terms result from the spatial dependent ellipticity. $C_k$ represents the coupling between the local and the traveling modes, it is small compared to the other terms so it will not be considered. $U_{km}$ and $V_{km}$ are scattering terms that describe the transition from a state with momentum  $\hbar k$  to another state with $\hbar m$. If we focus on the maximum scattering strength then the specific form of the coefficients, $U_{km}\sim\sech(k-m)$ and $V_{km}\sim\sech(k+m)$, allows us to approximate $U_{km}$ and $V_{km}$ by  delta functions $\delta_{km}$, $\delta_{k-m}$ respectively. We can then approximate $\delta H$ as
\begin{equation}\label{eqn:amplitudes}
\delta H=-c_{loc}^2\left(\frac{\pi D}{4\lambda}+K_\perp\right)+\sum_kA_kd_k^*d_k+B_k(d_k^*d_{-k}^*+d_kd_{-k}).
\end{equation}
The first term on the right hand side of Equation (\ref{eqn:amplitudes}) can be related to the domain wall mass by $p^2/(2m_N)\sim c_{loc}^2(\frac{\pi D}{4\lambda}+K_\perp)$, where $m_N=1/[2(\frac{\pi D}{4\lambda}+K_\perp)]$ is the N\'{e}el-type domain wall mass. \cite{hubertdomains, PSSB:PSSB19690320204}. The mass in a Bloch-type wall is $m_B=1/(2K_\perp)$ so $m_N<m_B$ which agrees with a higher mobility in DDWs.\cite{brataas_spintronics:_2013} The rest of the coefficients are
\begin{equation}
\begin{split}
A_k&=\omega_k(K_u+K_\perp)-\dfrac{K_\perp}{2}-\dfrac{\pi D}{4\lambda}\frac{(1+2k^2\lambda^2)}{\omega_k},\\
B_k&=\dfrac{K_\perp}{4}+\dfrac{\pi D}{8\lambda}\frac{(1+2k^2\lambda^2)}{\omega_k}.\\
\end{split}
\end{equation}
This Hamiltonian can be diagonalized by means of a Bogoliubov transformation, $c_k=u_k^+d_k-u_k^-d_k^*$,   $u_k^\pm = \sqrt{(A_k\pm\hbar\Omega_k)/2\hbar\Omega_k}$, to obtain
\begin{equation}
\delta H=-c_{loc}^2\left(\frac{\pi D}{4\lambda}+K_\perp\right)+\sum_k\hbar\Omega_kc_k^*c_k,
\end{equation}
where the frequency $\Omega_k$ is given by 
\begin{equation}
 \Omega_k=\dfrac{(K_u+K_\perp)a^3}{\hbar}\sqrt{\omega_k\left(\omega_k-\kappa-\dfrac{\pi D }{2\lambda(K_u+K_\perp)}\dfrac{(1+2k^2\lambda^2)}{(1+k^2\lambda^2)}\right)},
\end{equation}
%
where $a\sim0.3$ nm is the lattice constant. It is now possible to explicitly write the spin waves eigenmodes in terms of the amplitudes $c_k$, $c_{-k}^*$ and the local and traveling modes
\begin{equation}\label{eqn:swmode}
\chi(x)=i c_{loc}\xi_{loc}(x)+\sum_k(c_ku_k^++c_{-k}^*u_k^-)\xi_k(x)
\end{equation}
%
%
%
\section{Band structure in periodic wall arrays}
\begin{figure}
\centering\includegraphics[width=7.0cm]{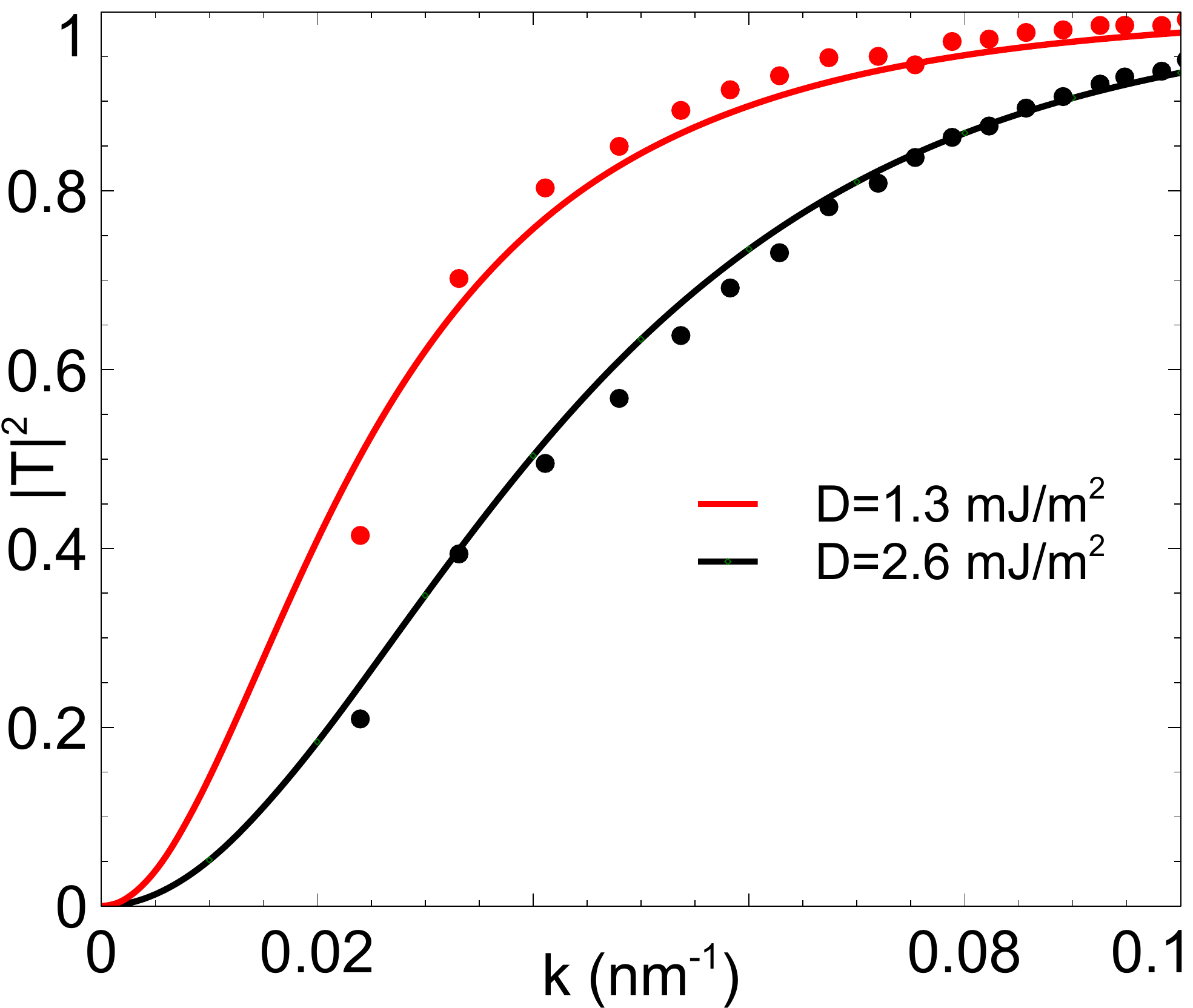}
\caption{(Color online) Transmission coefficient for $D=1.3$ mJ/m\textsuperscript{2} (red) and $D=2.6$ mJ/m\textsuperscript{2} (black). The solid lines result from using equation (\ref{depth}) and the points are  numerical simulations.}
\label{fig:coefficients}
\end{figure}

\begin{figure}
\centering\includegraphics[width=9.0cm]{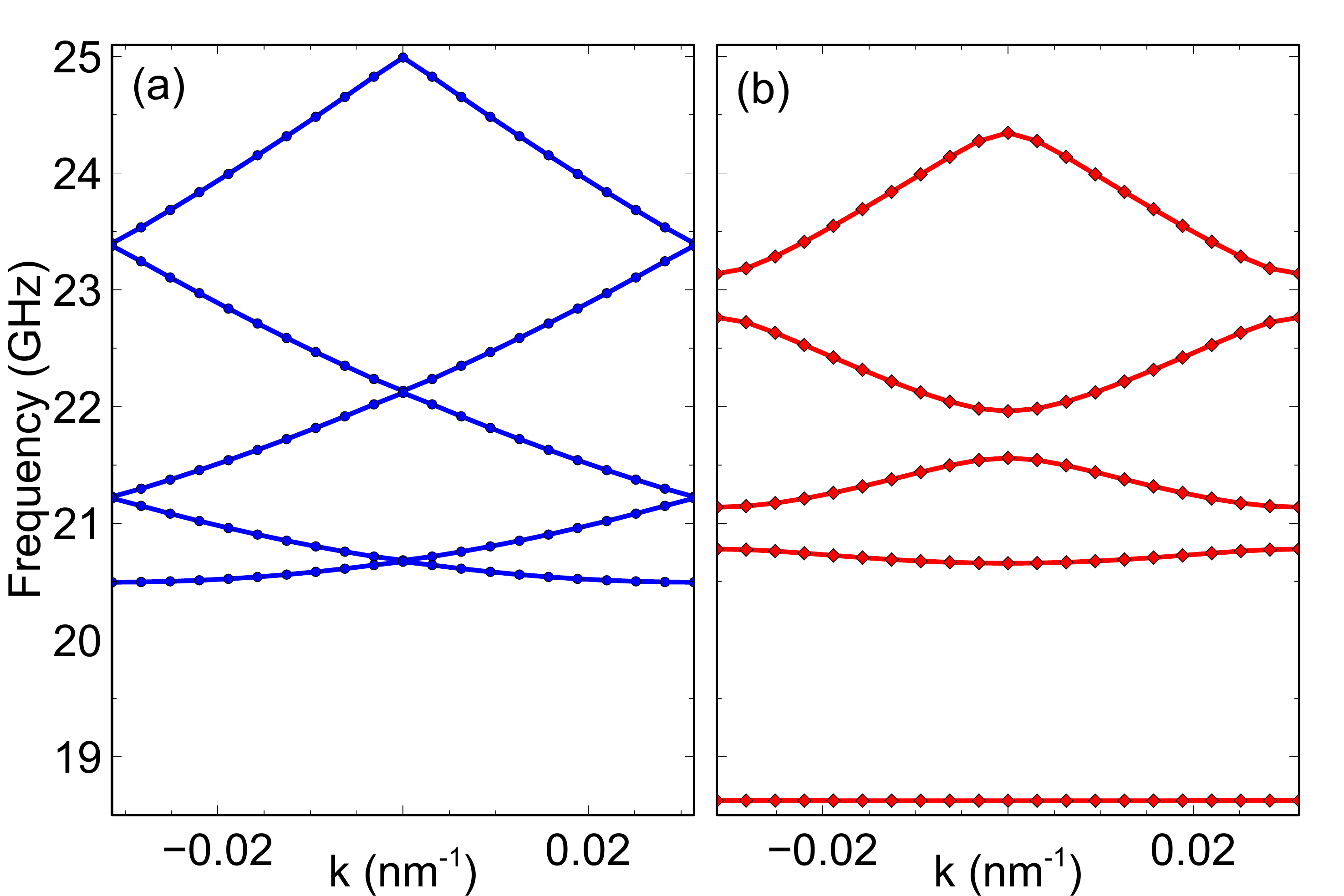}
\caption{(Color online)Band structures of a domain wall crystal. (a) Bloch wall ($D\,=\,0$).(b) N\'{e}el wall ($D=1.56$ mJ/m\textsuperscript{2}). }
\label{fig:bands}
\end{figure}

\begin{figure}
\centering\includegraphics[width=9.0cm]{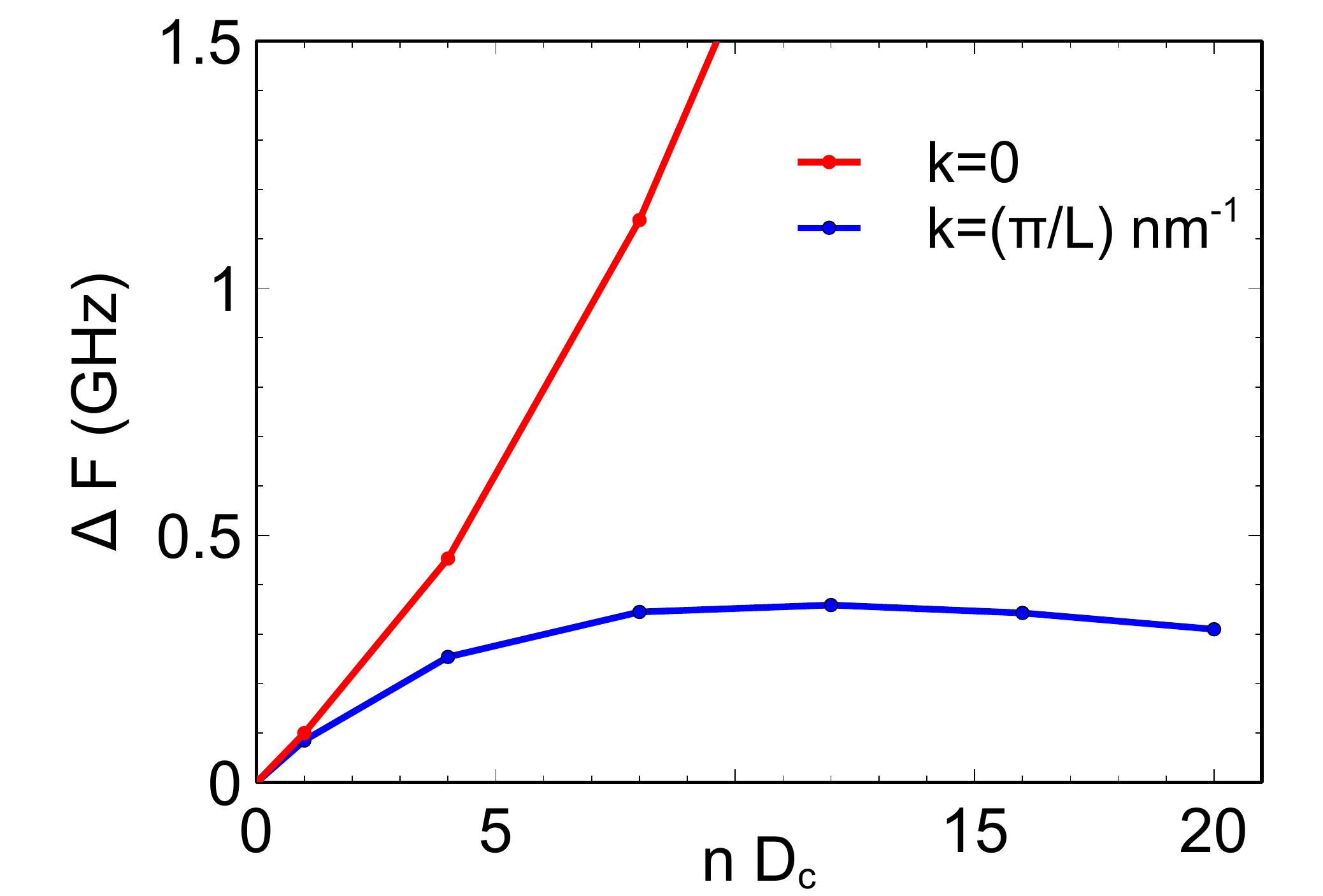}
\caption{(Color online) Frequency gaps $\Delta F$ at the Brillouin zone boundary as a function of $D_c$. $L=100$ nm\textsuperscript{-1} is the period of the crystal.}
\label{fig:gaps}
\end{figure}

The scattering potential for spin waves in a Bloch ($D=0$) domain  wall is represented by $V_P(x)$ which is reflectionless but leads to a a phase shift when spin waves  propagate through it. \cite{bayer_phase_2005, hertel_domain-wall_2004} $V_P(x)$ corresponds to a specific case of the so called modified P\"{o}schl-Teller Hamiltonian,\cite{lekner_reflectionless_2007}
\begin{equation}
\left[-\alpha^2\partial_x^2-l(l-1)\sech^2(x/\alpha)\right]\,\psi=\epsilon\,\psi.
\end{equation}
The parameter $l$ describes the depth of the potential well, $\alpha$ has units of distance and $\epsilon$ is a dimensionless energy. For a Bloch-type wall, $l=2$ and $\alpha=\lambda$. The transmission and reflection coefficients  related to the wave propagation across this potential have been calculated for this Hamiltonian as a function of the depth
\begin{equation}
|R|^2=\frac{1}{1+p^2};\;\;\;\;\;\;\;\;\;\;|T|^2=\frac{p^2}{1+p^2},\\
\end{equation}
with $p=\sinh(\pi k \alpha)/\sin(\pi l)$.\cite{flugge1971practical}  From this result it can be seen by inspection that for  $l\,\,\epsilon\,\,\mathbb{N}$,  $|R|^2$ is zero.
For the Dzyaloshinskii domain walls, the Hamiltonian is 
\begin{equation}\label{eqn:hamiltonian}
\left[-\lambda^2\partial_x^2-2\sech^2(x/\lambda)-\dfrac{D\sech(x/\lambda)}{\lambda(K_u+K_\perp)}\right]\chi(x)=E\chi(x),
\end{equation}
where the dimensionless energy is $E=\frac{\hbar\Omega_k}{(K_u+K_\perp)a^3}+\kappa-1$. 
The $D$ term modifies the depth of the potential but not its form. It possible then to relate the parameter $l$  with $D$, 
\begin{equation}\label{depth}
l=\frac{1}{2}\left[1+\sqrt{1+4\left(2+\dfrac{D}{\lambda(K_u+K_\perp)}\right)}\right].
\end{equation}
Two transmission coefficients were calculated as a function of the wave vector $k$ using Equation (\ref{depth})  for different values of $D$ and are shown in Figure (\ref{fig:coefficients}) along with numerical simulations to verify our theory. The numerical calculations were performed within a micromagnetic model. The calculations were done with the code mumax3.\cite{mumax3} The standard code includes the interface DMI term but was modified to include at the same time the in-plane and out-of-plane anisotropies. The parameters used were  $A=16$ pJ/m $K_\perp=18$ kJ/m\textsuperscript{3}, $K_u=0.5$ MJ/m\textsuperscript{3} and  $\lambda=5.55$ nm.  The system was discretized in cells of $1.5625\,\times\,1.5625\,\times\,1$ nm\textsuperscript{3}. The geometry coincides with the one showed in Figure 1 and the system size was $12800\,\times\,50\,\times\,1$ nm\textsuperscript{3} with periodic boundary condition in $y$ direction. To simplify the analysis and comparison with the analytical model the calculations were performed without damping term and demagnetizing field.  A domain wall was introduced at the center of the sample and then the system was excited with a monochromatic point source of $50$ mT applied field, $1950$ nm away from the domain wall. The amplitudes were calculated comparing the average envelope of the spin waves at both sides of the domain wall at the initial stages of the propagation. As $D$ increases significant reflection is found for larger values of $k$. This is a direct result of the scattering terms in Equation (\ref{eqn:ampl}). \\
As a result of the  DMI,  the scattering potential associated with the domain wall produces reflection in the spin waves propagating through it. As such, the band structure for a lattice of DDWs  presents gaps at the edges of the Brillouin zone because of Bragg reflection, which is  not present for Bloch-type walls for which no reflection occurs.
To see this we consider a periodic array of DDWs and Fourier transform Equation (\ref{eqn:hamiltonian}) using also Bloch's theorem on $\chi(x)$ to obtain the central Equation 
\begin{equation}\label{eqn:centraleqn}
\left((K_u+K_\perp)\lambda^2k^2-E\right)C(k)+\sum_GU_GC(k-G)=0,
\end{equation}
where $U_G$ are the Fourier coefficients of the potential.\cite{kittel2004introduction} The period of the DDW crystal can be determined with the Kooy-Enz formula that describes the stray field energy for an arrangement of parallel band domains separated by domain walls of zero width.\cite{kooy1960experimental} For a particular case of $D=2.6$ mJ/m\textsuperscript{2} a minimum film thickness  of approximately $2$ nm is found with a period of $L=100$ nm. There is a compromise between the film thickness and the period, since the minimum film thickness and period increase as $D$ decreases.\\ 
Equation (\ref{eqn:centraleqn}) represents an infinite set of equations connecting the coefficients $C(k-G)$ for all reciprocal lattice vectors $G$. These equations are consistent if the determinant of the coefficients is zero. It is often only necessary to consider the determinant of a few coefficients. For our calculations an $11\times11$ matrix is used to numerically solve the central equation.\\
The calculated band structure of domain wall crystals is shown in figure~\ref{fig:bands}. Gaps in the band structure are a consequence of Bragg reflection and a direct result of the DMI. Figure (\ref{fig:gaps}) shows the first gap frequencies as a function of $D$ in $k=0$ and $k=\pi/L$. 
%
%
%
\section{Discussion and Concluding Remarks}
Reflection of spin waves by a domain wall is found when the interface form of DMI is included. It is a result of the stabilization of a N\'{e}el wall as the stable configuration, and of an extra chiral term in the Hamiltonian that changes the P\"{o}schl-Teller potential and scatters the spin waves. Reflection results in energy gaps in the band structure of a periodic array of domain similar to the ones found in a magnonic crystal. Our proposed model offers an alternative method for constructing a magnonic crystal without the need to build the metamaterial although we recognize the difficulty of stabilizing the domains. Moreover, the gaps only depend on $D$ so there is only one parameter to control. It is worth noting that the bulk form of DMI favors a Bloch-type wall configuration and no extra term is found in the spin wave energy. This last statement agrees with previous claims that the reflectionless feature is very robust.\cite{yan_all-magnonic_2011}
%

%
%
This work was partially supported by the University of Glasgow, EPSRC (EPSRC EP/M024423/1\cite{doi}) , the National Council of Science and Technology of Mexico (CONACyT), and the French National Research Agency (ANR) under contract no. ANR-11-BS10-003 (NanoSWITI).
%
\newpage
\section{References}
\bibliography{magcrys1.bib}
\end{document}